\documentclass[pra,superscriptaddress,nofootinbib,
 amsmath,amssymb, aps, twocolumn]{revtex4-2}
\usepackage{xr-hyper}
\makeatletter
\usepackage[unicode=true,pdfusetitle,hypertexnames=false,
 bookmarks=true,bookmarksnumbered=false,bookmarksopen=false,
 breaklinks=true,pdfborder={0 0 0},pdfborderstyle={},backref=false,colorlinks=true]
 {hyperref}
\usepackage[justification=justified, format=plain]{caption}
\usepackage{braket}
\usepackage{svg}
\usepackage{float}
\usepackage{caption}
\usepackage{tikz}
\usepackage{quantikz}
\usepackage{subcaption}
\usepackage{placeins} 
\usepackage{graphicx}
\usepackage{dcolumn}
\usepackage{bm}

\DeclareMathOperator*{\argmin}{arg\,min}

\begin{document}

\title{Compact fermionic quantum state preparation \\ with a natural-orbitalizing variational quantum eigensolving scheme}

\author{P. Besserve}
\affiliation{Eviden Quantum Laboratory, 78340 Les Clayes-sous-Bois, France
}
\affiliation{CPHT, CNRS, Ecole Polytechnique, Institut Polytechnique de Paris, 91128 Palaiseau, France}
\author{M. Ferrero}
\affiliation{CPHT, CNRS, Ecole Polytechnique, Institut Polytechnique de Paris, 91128 Palaiseau, France}
\affiliation{Coll\`ege de France, 11 place Marcelin Berthelot, 75005 Paris, France}
\author{T. Ayral}
\affiliation{Eviden Quantum Laboratory, 78340 Les Clayes-sous-Bois, France
}

\begin{abstract}
Assemblies of strongly interacting fermions, whether in a condensed-matter or a quantum chemistry context, range amongst the most promising candidate systems for which quantum computing platforms could provide an advantage. 
Near-term quantum state preparation is typically realized by means of the variational quantum eigensolver (VQE) algorithm.
One of the main challenges to a successful implementation of VQE lies in the sensitivity to noise exhibited by deep variational circuits.
On the other hand, sufficient depth must be allowed to be able to reach a good approximation to the target state.
In this work, we present a refined VQE scheme that consists in topping VQE with state-informed updates of the elementary fermionic modes (spin-orbitals). These updates consist in moving to the natural-orbital basis of the current, converged variational state, a basis we argue eases the task of state preparation. 
We test the method on the Hubbard model in the presence of experimentally relevant noise levels.
For a fixed circuit structure, the method is shown to enhance the capabilities of the circuit to reach a state close to the target state without incurring too much overhead from shot noise. Moreover, coupled with an adaptive VQE scheme that constructs the circuit on the fly, we evidence reduced requirements on the depth of the circuit as the orbitals get updated.
\end{abstract}

\maketitle

Many-body systems such as strongly interacting condensed-matter systems and chemical compounds are expected to range amongst the first systems whose understanding could benefit from quantum computational power \cite{epja_review}.
Their description with classical computers indeed always faces an exponential wall: it may stem from the exponential size of the Hilbert space with the number $N_e$ of particles, which strongly limits the scope of exact diagonalization techniques (also known as full configuration interaction in chemistry), or an exponentially growing variance in Monte Carlo methods due to the fermionic sign problem, or an exponential bond dimension in tensor-network methods in volume-law regimes.

The first prototypes of quantum computers could help overcome this exponential barrier. 
Current prototypes, dubbed noisy, intermediate scale quantum (NISQ, \cite{preskill_quantum_2018}) computers, feature high noise levels and low qubit counts. In the absence of quantum error correction, the fidelity of quantum states decreases exponentially with the circuit depth, placing very strong limitations on the number of gates that can be executed reliably \cite{Louvet2023}. 
This precludes the use of "textbook" quantum algorithms with provable speedups such as quantum phase estimation, and ask for the development of algorithms that require only very shallow circuits. This is typically achieved by doing most of the heavy-lifting with classical computers (CPUs) and only spoon-feeding very specific tasks to the quantum processing unit (QPU).

The epitome of such hybrid algorithms is the variational quantum eigensolver (VQE \cite{peruzzo_variational_2014}, see \cite{cerezo_variational_2020, tilly_variational_2021} for a review): 
one defines a parameterized circuit (also called ansatz) and tunes its parameters by minimizing the energy, a quantum observable over the circuit instance, with a classical optimizer. 
The success of the (classical) parameter optimization is by no means guaranteed due to the barren plateau problem (\cite{mcclean_barren_2018}, see \cite{Larocca2024} for a review), which colloquially stems, as the number of qubits gets large, from (i) a variational circuit which is too expressive, or (ii) a level of noise which is too high, or (iii) an observable which is too global, or (iv) an  initial state that is too entangled \cite{Ragone2023}.

\begin{figure}
     \centering
    \includegraphics[width=\columnwidth]{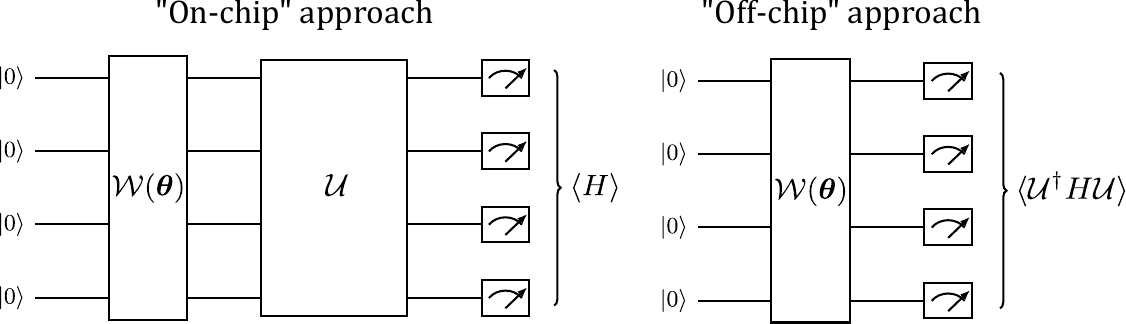}
    \caption{"On-chip" (purely quantum, left) vs "off-chip" (classical, right) strategy to rotate the single-particle basis of a variational ansatz $\mathcal{W}(\bm{\theta})$.}
    \label{fig:onchip_offchip}
\end{figure}

In this work, we address issues (i) and (ii) by focusing on a degree of freedom specific to fermionic problems, namely the single-particle orbital basis from which fermionic many-body states are constructed.
There are two ways to perform a single-particle orbital basis rotation $\mathcal{U}^{\dagger} H \mathcal{U}$ on a Hamiltonian $H$, illustrated in Fig.~\ref{fig:onchip_offchip}.
On the one hand, one can rotate, using so-called Givens rotations \cite{wecker_solving_2015}, quantum states using a quantum circuit \emph{on a quantum computer}. This approach has been carried out e.g in \cite{arute_hartree-fock_2020} and is the inspiration of some fermionic ansätze \cite{dallaire-demers_low-depth_2019}.
On the other hand, one can change the representation of the Hamiltonian by performing $\mathcal{U}^{\dagger} H \mathcal{U}$ \emph{on a classical computer}: since $\mathcal{U}$ corresponds to a single-particle basis change, the corresponding computation is classically efficient (i.e polynomial in the number of orbitals). We will dub this approach the "off-chip" approach as it is not performed on the quantum chip.
Deciding on one or the other strategy results from the following tradeoff: 
performing the basis rotations on the quantum computer will render the computation more sensitive to quantum noise, but can reduce the number of terms in the Hamiltonian and thus the measurement overhead \cite{Huggins2019,Yen2021}.
Conversely, performing the basis change classically at the level of the Hamiltonian could increase the number of terms and thus the measurement overhead, but also help reduce the circuit depth and thus make it more noise-robust.

Here, we focus on the second, "off-chip" approach because we are focusing on current (noisy) devices.
Several such orbital optimization schemes have been proposed in the past few years in various VQE or broader variational contexts: mainly quantum chemistry \cite{koridon_orbital_2021, sokolov_quantum_2020, mizukami_orbital_2020, ratini_wave_2022, Bierman2022, Fitzpatrick2022}, but also nuclear physics    \cite{robin_quantum_2023}, tensor networks \cite{Zgid2008, Ghosh2008} as well as neural networks \cite{moreno_enhancing_2023}.
In these schemes, $\mathcal{U}$ is optimized through parameters that are put on an equal footing with that of the variational circuit.
These approaches either add a large number of parameters to optimize, and/or a large number of gradients to be evaluated in the case of adaptive ansatz construction \cite{grimsley_adaptive_2019}.

Here, we present a lighter orbital-adapting technique consisting in looking for a physically-motivated orbital basis instead of brute-force search of an optimal basis. More specifically, we propose to leverage the so-called \textit{natural-orbital} (NO) basis, which has been extensively used for quantum chemistry in a classical computing context. 
 
We propose two different versions of our natural-orbitalization strategy.
The first version (already introduced by some of us in the context of Anderson impurity models, \cite{besserve_unraveling_2022}) consists in picking an ansatz circuit and simply interlacing plain-vanilla VQE runs with rotations to the optimal state's NO basis.
Doing so, we observe, for the Hubbard model, considerable increases in the expressiveness of the ansatz at hand, rendering very shallow---and thus, noise-resilient---ansätze capable of preparing the target state.
For deep circuits, noisy performances are also greatly improved.
Interestingly, we find that the increase in the number of Hamiltonian Pauli terms incurred by rotating the single-particle basis seems to have a limited impact on the measurement overhead due to the concentration of Pauli weights around a few terms only.

The second version consists in combining the natural-orbitalization framework with the ADAPT-VQE method \cite{grimsley_adapt-vqe_2022} which constructs the circuit iteratively. This allows to both 'learn' a circuit and a representation. In this framework, we observe that very shallow circuit ansätze can be discovered even in the presence of noise.

\section{Variational quantum eigensolving with fermions: the role of the orbital basis}

\subsection{Fermions and the orbital basis}
We are interested finding the ground-state energy of a fermionic Hamiltonian $H$ given in its second-quantized form
\begin{equation}
        H = \sum \limits_{p,q=1}^M h_{pq}c^{\dagger}_p c_q + \frac{1}{2}\sum \limits_{p,q,r,s=1}^M h_{pqrs}c^{\dagger}_p c^{\dagger}_q c_r c_s,
\end{equation}
with $h_{pq}$ and $h_{pqrs}$ the kinetic and interaction tensors, and $c^{\dagger}_p, c_p$ the creation and annihilation operators for the $p$-th spin-orbital ($p$ encompasses both a spatial and a spin degree of freedom).

Any many-body state $\ket{\psi}$ (including the ground state of $H$) can be written as a linear combination
\begin{equation}
    \ket{\psi} =\sum_{n_1 n_2\dots n_M \in \{0, 1\}^M} \psi_{n_1 n_2 \dots n_M} \ket{\phi_{n_1, \dots, n_M}}
    \label{eq:Fock_decomp}
\end{equation}
of an exponential number of Fock states
\begin{equation}
    \ket{\phi_{n_1, \dots, n_M}} = (c^{\dagger}_1)^{n_1}\cdots(c^{\dagger}_M)^{n_M}  \ket{\Bar{0}},
\end{equation}
which are multi-fermion states corresponding to the excitation of some of the fermionic modes through creation operators $c^{\dagger}_p$ over the fermion vacuum state $\ket{\Bar{0}}$.
There exists a freedom in the choice of this basis of single-particle fermionic modes $c^{\dagger}_p$, namely one could also write 
\begin{equation}
    \ket{\psi} =\sum_{n_1 n_2\dots n_M \in \{0, 1\}^M} \tilde{\psi}_{n_1 n_2 \dots n_M} \ket{\tilde{\phi}_{n_1, \dots, n_M}}
\end{equation}
with Fock states $\ket{\tilde{\phi}_{n_1, \dots, n_M}} = (\tilde{c}^{\dagger}_1)^{n_1}\cdots(\tilde{c}^{\dagger}_M)^{n_M}  \ket{\Bar{0}}$ built out of rotated fermionic modes:
\begin{equation}
    \tilde{c}^{\dagger}_i = \sum_p R_{pi}c^{\dagger}_p, \;\;\;
    \tilde{c}_i = \sum_p R^{\dagger}_{ip} c_p,
\end{equation}
with $R \in U(M)$ to ensure the preservation of the anticommutation rules.

The number of nonzero coefficients $\psi_\alpha$ (with $\alpha = n_1, \dots, n_M$) in the first representation is in general different from the number of nonzero coefficients $\tilde{\psi}_\beta$.
An extreme case is the case when $|\psi_0\rangle$ can be represented as a single Fock state in a basis (e.g the $c^\dagger_p, c_p$ basis): only one $\psi_\alpha$ coefficient is nonzero.
The representation of such a state, called a Slater determinant, can have many nonzero $\tilde{\psi}_\beta$ coefficients in another basis.

The rotation $R$ will lead to the following equivalent expression for the Hamiltonian:
\begin{equation}
        H = \sum \limits_{i,j=1}^M \tilde{h}_{ij}\tilde{c}^{\dagger}_i \tilde{c}_j + \frac{1}{2}\sum \limits_{i,j,k,l=1}^M \tilde{h}_{ijkl}\tilde{c}^{\dagger}_i \tilde{c}^{\dagger}_j \tilde{c}_k \tilde{c}_l,
        \label{eq:H_rotated}
\end{equation}
with modified tensors 
\begin{subequations}
\begin{align}
    \tilde{h}_{ij} &= \sum_{pq} R^\dagger_{ip} h_{pq} R_{q j}, \\
    \tilde{h}_{ijkl} &= \sum_{pqrs}
    R^\dagger_{ip} R^\dagger_{jq} h_{pqrs} R_{rk} R_{sl}.
\end{align}
\end{subequations}

One can thus seek to optimize the rotation basis $R$. For instance, one may want to reduce the number of nonzero coefficients of the wavefunction, or, alternatively, reduce the number of terms in the Hamiltonian (Eq. (\ref{eq:H_rotated})).

Before we turn to the optimization \textit{per se}, 
let us note that a change in $R$ will effect a change of coordinates
 \begin{equation}
     \tilde{\psi}_\beta = \sum_\alpha \mathcal{U}_{\beta \alpha} \psi_\alpha,
 \end{equation}
 with the expression of the many-body unitary transformation $\mathcal{U} \in U(2^M)$ as a function of $R$ given by Thouless's theorem \cite{thouless_stability_1960} 
 \begin{equation}
 \label{eq:thouless}
 \mathcal{U} = e^{-\sum_{pq} \left(\log R\right)_{pq}c^{\dagger}_p c_q}.
 \end{equation}

 In other words, an orbital change of basis is a unitary operation $\mathcal{U}$ on the many-body wavefunction with the specific form given in Eq.~\eqref{eq:thouless}.

\subsection{Optimizing the single-particle basis} \label{subsec:optimizing_orbital_basis}

In a standard VQE workflow, one seeks to optimize the variational energy
\begin{equation}
    E(\bm{\theta}) = \langle \Bar{0} | \mathcal{W}(\bm{\theta})^\dagger H  \mathcal{W}(\bm{\theta}) |\Bar{0} \rangle,
\end{equation}
with $\mathcal{W}(\bm{\theta})$ a parameterized unitary implemented with quantum gates.
Adding an orbital optimization amounts to replacing the energy function by 
\begin{equation}
    E(\bm{\theta}, \bm{\theta'}) = \langle \Bar{0} | \mathcal{W}(\bm{\theta})^\dagger \mathcal{U}(\bm{\theta'})^\dagger  H \mathcal{U}(\bm{\theta'})  \mathcal{W}(\bm{\theta}) | \Bar{0} \rangle,
\end{equation}
with $\bm{\theta'}$ a vector of the $M(M-1)/2$ parameters defining the anti-hermitian matrix $\log R$ introduced in Eq.~\eqref{eq:thouless}. 

Many VQE schemes exploiting this freedom in the choice of single-particle modes have been proposed in the past few years. They differ in (i) whether the rotation is done classically ("off-chip", at the level of the Hamiltonian) or quantumly ("on-chip", at the level of the state), as illustrated in Fig.~\ref{fig:onchip_offchip}, and (ii) what they attempt to optimize with this new degree of freedom.

"Off-chip" schemes \cite{mizukami_orbital_2020, sokolov_quantum_2020, ratini_wave_2022, moreno_enhancing_2023,  Fitzpatrick2022, Bierman2022, robin_quantum_2023} consist in dressing the Hamiltonian on a classical computer
\begin{equation}
\label{eq:dressed_H}
    \tilde{H}(\bm{\theta'}) = \mathcal{U}(\bm{\theta'})^{\dagger} H\mathcal{U}(\bm{\theta'}). 
\end{equation}
and preparing the wavefunction $\mathcal{W}(\bm{\theta}) | 0 \rangle$ on a quantum computer. This is a classically efficient (polynomial) operation thanks to Thouless's theorem.

One optimization criterion is to
minimize of the one-norm of the Hamiltonian \cite{koridon_orbital_2021}: after decomposing $\tilde{H}(\bm{\theta'})$ in terms of Pauli words $\tilde{H}(\bm{\theta'}) = \sum_d \lambda_d H_d$ 
given an encoding, its one-norm is defined as $\left\Vert \tilde{H}(\bm{\theta'})\right\Vert_1 = \sum_d |\lambda_d |$.
The standard error $\Delta E$ stemming from a limited number $n_{\mathrm{shots}}$ of shots is upper-bounded by $\left\Vert \tilde{H}(\bm{\theta'})\right\Vert_1 / \sqrt{n_{\mathrm{shots}}}$ \cite{Rubin2018}, so that minimizing the one-norm should decrease the measurement overhead for precise evaluation of the energy in quantum chemistry.

Another more widespread optimization criterion is to minimize the energy with the help of these additional parameters $\bm{\theta'}$.
In \cite{mizukami_orbital_2020, moreno_enhancing_2023, robin_quantum_2023}, the optimization on $\bm{\theta}$ and $\bm{\theta'}$ is done jointly, while it is done in an alternate fashion in  \cite{ratini_wave_2022}.

The basis in which the Hamiltonian is written---namely the very form of $\tilde{H}(\bm{\theta'})$---generally impacts the behavior of the optimization.
For instance, in \cite{ratini_wave_2022}, greater correlation energy was observed for molecular systems such as $LiH$ with shallow empirical ansätze when carrying out orbital optimization along VQE.
Along with boosted expressiveness, this could be due to a favorable effect of the variational unitary dressing onto the optimization landscape \cite{moreno_enhancing_2023}.

An extreme example of the off-chip vs on-chip tradeoff and of the off-loading of the optimization difficulty to classical vs quantum processors is supplied by the implementation of the Hartree-Fock method presented in \cite{arute_hartree-fock_2020}.
In this work, the Hartree-Fock method, which consists in finding $R$, or equivalently $\mathcal{U}(\bm{\theta'})$, that minimizes the energy (i.e here $\mathcal{W}(\bm{\theta})=I$), is implemented entirely "on-chip": states $\mathcal{U}(\bm{\theta'})|0\dots 0,1\dots 1\rangle$ are prepared on a superconducting processor with so-called Givens rotation and their energy measured.
By contrast, the usual implementation of the Hartree-Fock method can be regarded as performing the orbital rotation "off-chip": in this case, the "quantum" part, namely the state preparation, is trivial as it consists in producing states $|0\dots 0,1\dots 1\rangle$.
This extreme example illustrates the fact that if the Hamiltonian is adequately (classically) preprocessed, the depth required to reach its ground state may be reduced.

One major drawback of the above methods is that they incur a rather large cost in terms of classical computation: the parameter space is enlarged, and gradient-based minimizations will require more circuit samples than the mere circuit parameter update does.
Additionally, more complex landscapes arising from this added variational freedom have been reported on for chemical systems studied with the unitary coupled-cluster (UCC) ansatz \cite{sokolov_quantum_2020}. 

This motivates our choice to use a physically-motivated orbital change instead of optimizing a parameterized orbital change. 
Here, our target orbital change will be the so-called \textit{natural orbitals} (NO).

\subsection{Natural orbitals}

Natural orbitals \cite{lowdin_quantum_1955} are defined as the single-particle orbitals that diagonalize the \textit{one-particle reduced density matrix} (1-RDM) $D$, defined by its matrix elements
\begin{equation}
    \label{eqn:def_1RDM}
    D_{pq} \equiv \langle \psi |c^{\dagger}_p c_q |\psi \rangle,
\end{equation}
of a given $M$-qubit fermionic quantum state  $\ket{\psi}$.
Namely, if $D$, a positive semidefinite matrix, is diagonalized by the unitary transformation $V$, $D = V \mathrm{diag}(n_1 \dots n_M) V^\dagger$, then the creation/annihilation operators  
\begin{equation}
     \tilde{c}^{\dagger}_i = \sum_p V^{\dagger}_{ip}c^{\dagger}_p, \;\;\;
    \tilde{c}_i = \sum_p V_{pi} c_p,
\end{equation}
create/annihilate electrons in the $i$-th \textit{natural orbital}. The diagonal elements $n_1 \dots n_M$  are called the natural orbital occupation numbers (NOONs) and are such that $0\leq n_k \leq 1$, 0 meaning an empty orbital and 1 an occupied orbital.

The natural-orbital basis is colloquially said \cite{lowdin_quantum_1955} to be the basis that minimizes the number of Fock states (or Slater determinants) in the decomposition of a given state (or equivalently minimizes the number of nonzero coefficients in Eq.~\eqref{eq:Fock_decomp}).
A more rigorous statement (proven in Appendix \ref{app:proof_s_corr_min}) is that the NO basis minimizes the correlation entropy
\begin{equation}
    S_{\mathrm{corr}} = - \sum_p D_{pp} \log{D_{pp}}
\end{equation}
where $D_{pp}$ is the occupancy of the $p$-th single-particle orbital.
Thus, put informally, the NO basis has the least partially filled orbitals, which explains the original colloquial statement (since states with no partial occupations are single Slater determinant, aka single-reference states, states with the least partial occupations should be representable by the fewest Slater determinants).

This property has prompted the widespread use of NOs in classical computational methods.
In quantum chemistry, configuration interaction (CI) expansions based on natural orbitals were found to range amongst the fastest and most accurate (see e.g. \cite{lowdin_quantum_1955, davidson_natural_1972}).
NOs have also been used to tackle Anderson impurity models \cite{lin_efficient_2013, lu_exact_2017, lu_natural-orbital_2019, bi_natural_2019}. There, they provide a way to alleviate the effect of the harsh bath truncations that the exponential scaling of the Hilbert space with the number of bath sites calls for.

We claim that the properties of the natural orbitals can also be beneficial for the preparation of quantum states with quantum processors:
intuitively, states that can be represented as a linear combination of few Slater determinants will require less complex circuits than states requiring more Slater determinants. 
In a way, the NO basis should be the basis such that single-particle orbital rotations (with generators quadratic in the creation and annihilation operators, such as in Eq.~\eqref{eq:thouless}) have been stripped away from the circuit and put inside the representation of the Hamiltonian, so that the quantum processor only executes "nontrivial" unitary operations.

Of course, we have overlooked one major hurdle so far: computing the NO transformation requires the knowledge of the 1-RDM $D$ and thus of the state $\ket{\psi}$ it is defined upon... Yet, this state is also the ground state we are looking for. 

In the next section, we will present an iterative method to solve this conundrum.

Before this, let us mention recent work \cite{Ratini2023a} implementing the kind of orbital optimization we mentioned in subsection \ref{subsec:optimizing_orbital_basis} and that a posteriori backs up our choice: in this work, using heuristic ladder ansätze, the orbital optimization scheme was observed to converge towards orbitals very close to the NOs for most of the chemical compounds investigated ($LiH$, $H_2$, $H_2O$, $HF$, $NH_3$).

\section{Methods}

\begin{figure}
    \centering
    \includegraphics[width=\columnwidth]{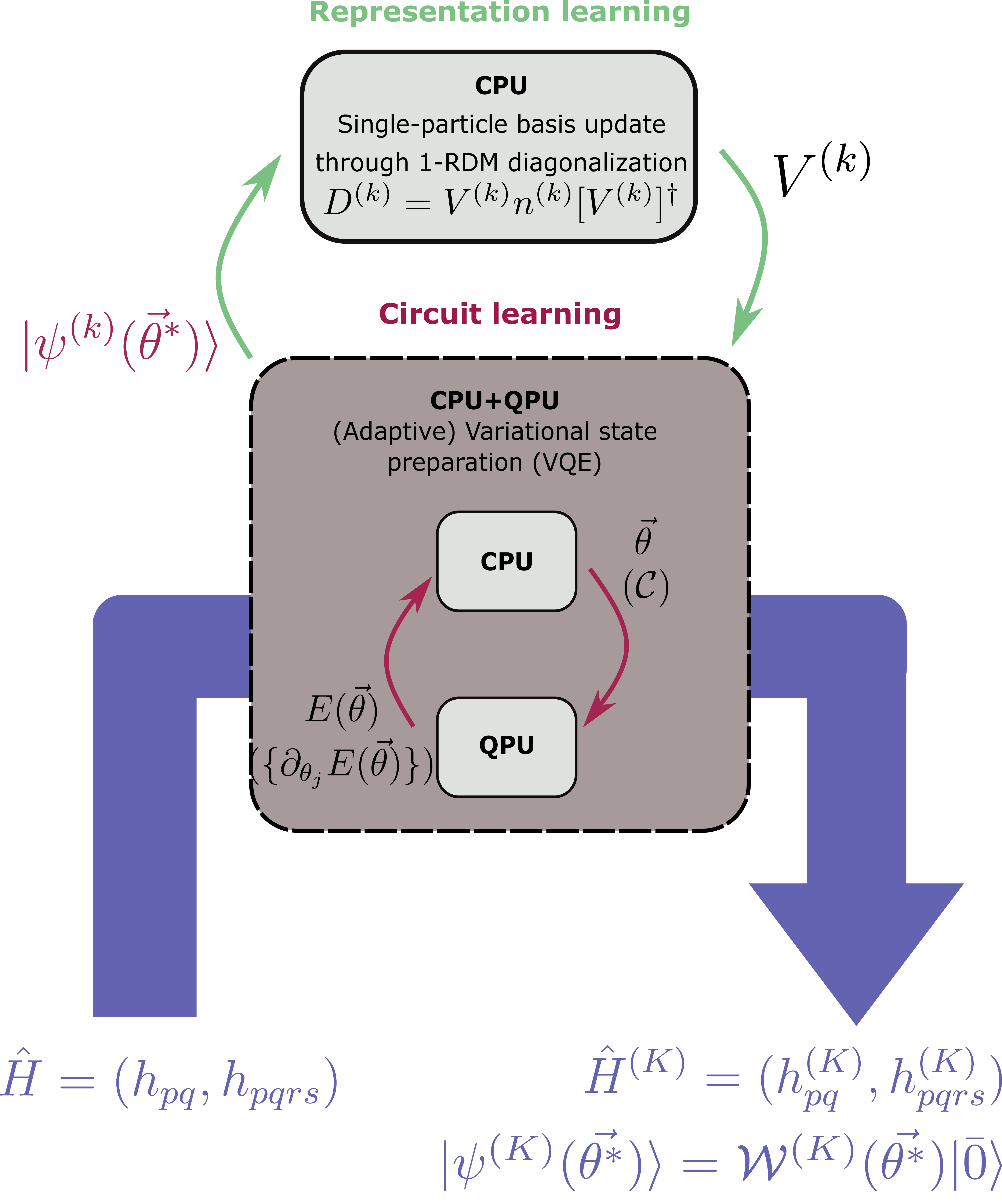}
    \caption{\emph{Flow diagram of the natural-orbitalizing VQE methods}. The algorithm receives a Hamiltonian $\hat{H}$, defined by the tensors $h_{pq}$ and $h_{pqrs}$, whose ground state is to be determined. Repeating interleaved sequences of (possibly adaptive) VQE runs and single-particle basis updates until convergence, the algorithm outputs a final variational approximation $\ket{\psi^{(K)}}=\mathcal{W}^{(K)}(\bm{\theta}^*) \ket{\Bar{0}}$ to the target ground state, expressed in the rotated basis associated with the transformed Hamiltonian $\hat{H}^{(K)} = (h^{(K)}_{pq}, h^{(K)}_{pqrs})$.}
    \label{fig:method}
\end{figure}

As sketched in the previous section, the natural orbital basis we are aiming to rotate our representation into requires the prior knowledge of the very ground state we are looking for.
In this section, we describe how we solve this problem. 
Essentially, we gather information about the true NO basis by computing the NO basis corresponding to the current converged VQE state, in the hope that as the basis gets rotated to a basis closer to the true NO basis, the quality of VQE, and thus of the converged state, improves.

In other words, we propose to interlace VQE runs---usual "circuit learning" steps---with rotations onto the natural-orbitals basis of the current variational state---adding a representation learning aspect, as illustrated on Fig.~\ref{fig:method}.
This amounts to pushing the QPU-CPU hybridization further. 
We dub this method 'natural-orbitalization' and abbreviate it as NOization \cite{besserve_unraveling_2022}.

\subsection{Representation learning through natural-orbitalization (NOization)}
In the first variant of the algorithm, introduced by some of us to solve Anderson impurity models \cite{besserve_unraveling_2022}, a variational circuit is picked once and for all, as in usual VQE.
A first VQE optimization over the circuit parameters in the original basis selected to write down the Hamiltonian initially (typically, in a condensed-matter context, the site-spin basis) is run in order to provide a first approximation to the ground state.
The 1-RDM of the attained state is then diagonalized to provide a new set of orbitals to work with. The VQE procedure is then repeated, and the orbitals get re-updated according to the new converged state.
The number of orbital rotations is a hyperparameter of the method we note $K$. We will see that a few steps ($K \simeq 3$) are enough to drastically increase the expressiveness of the circuit. 

Let us now spell out how to perform the orbital rotation. At step $k \in 0\dots K - 1$ of the method, one aims at preparing the ground state of the current Hamiltonian
\begin{equation}
    H^{(k)} = \sum \limits_{pq} h^{(k)}_{pq}c^{\dagger}_p c_q + \frac{1}{2}\sum \limits_{pqrs} h^{(k)}_{pqrs}c^{\dagger}_p c^{\dagger}_q c_r c_s. 
\end{equation}
The tensors $h^{(k)}_{pq}$ and $h^{(k)}_{pqrs}$ are initialized e.g. as the usual $h_{pq}$ and $h_{pqrs}$ tensors of the Hamiltonian in the site-spin basis. They are the objects which will be rewritten to reflect the single-particle orbital rotation.

The key step of the algorithm goes as follows. Provided the optimal VQE state $\ket{\psi(\bm{\theta}^{*(k)})}$ returned by VQE at NOization step $k$ of the procedure---namely minimizing the expectation value of $H^{(k)}$---we use the quantum computer to compute the matrix elements of the 1-RDM 
\begin{equation}
    D^{(k)}_{pq} = \langle \psi(\bm{\theta}^{*(k)}) | c^{\dagger}_p c_q|\psi(\bm{\theta}^{*(k)}) \rangle.
\end{equation}
In the presence of noise, the VQE optimization yields a density matrix $\rho(\bm{\theta}^{*(k)})$ instead of a pure state $\ket{\psi(\bm{\theta}^{*(k)})}$ and, accordingly, we actually measure expectation values that read
\begin{equation}
    D^{(k)}_{pq} = \mathrm{Tr}(\rho(\bm{\theta}^{*(k)})c^{\dagger}_p c_q).
\end{equation}
Note that this quantity can actually be reconstructed at no additional cost by aggregating the corresponding terms measured to compute the energy of the variational state $|\psi(\bm{\theta}^{*(k)}) \rangle$ (resp. $\rho(\bm{\theta}^{*(k)})$). Then, we classically compute the transformation $V^{(k)}$ that diagonalizes $D^{(k)}$:
\begin{equation}
D^{(k)}_{pq} = \sum \limits_{\alpha} V^{(k)}_{p\alpha} n_{\alpha} V^{(k)\dagger}_{\alpha q}.
\end{equation}
This matrix is used to update the orbital basis as:
\begin{equation}
    c_{\alpha} \rightarrow  \sum_q
    V^{(k)}_{q \alpha} c_q
\end{equation}
which effectively corresponds to the following transformation being applied onto the fermionic Hamiltonian's coefficients:
\begin{subequations}
\begin{align}
    &h^{(k+1)}_{pq}= \sum \limits_{p'q'} V^{(k)}_{p'p}h^{(k)}_{p'q'}(V^{(k)\dagger})_{qq'} \\
    &h^{(k+1)}_{pqrs}= \sum \limits_{p'q'r's'}V^{(k)}_{p'p}V^{(k)}_{q'q}h^{(k)}_{p'q'r's'}(V^{(k)\dagger})_{rr'}(V^{(k)\dagger})_{ss'}.
\end{align}
\end{subequations}
The procedure is repeated $K$ times until a convergence criterion is met or an update budget is reached. As will be evidenced in the Results section, the effect is to boost the expressiveness of the quantum circuit.

The strategy described above is similar in spirit to that implemented in the so-called Perm-VQE method \cite{tkachenko_correlation-informed_2021}, in which qubit ordering (as opposed to the full single-particle orbital basis) is updated iteratively, so that leading correlations are displayed by neighboring qubits.

\subsection{Representation \textit{and} circuit learning: natural-orbitalizing adaptive VQE (NOA-VQE) scheme}
\label{sec:NOA-VQE}
The previous method works with a fixed circuit structure, and aims at enhancing the expressiveness of that circuit.

Having a fixed ansatz imposes some limitations. 
For instance, in the noninteracting limit, $h_{pqrs}=0$, we know that the ground state is a single Fock state in the NO basis.
Therefore, in the NO basis, a circuit with only single-qubit rotations will capture the solution. But in order to reach this NO basis via NOization, we need to start from the original basis, in which the ground state is not a single Fock state... and therefore requires entangling circuits. 
We thus need to allow for an adaptive ansatz to capture this limit.
In fact, this strategy is probably useful in other limits: not only should the ansatz adapt to the current single-orbital basis, but also to the noise level.

We propose to resort to ADAPT-VQE \cite{grimsley_adaptive_2019} (in its "QUBIT" version \cite{tang_qubit-adapt-vqe_2021}) to construct the circuit at step $k$ of the NOization procedure.
This variant simply consists in using ADAPT-VQE in the VQE box of Fig.~\ref{fig:method} instead of plain VQE. By allowing the circuit to be constructed on the fly, one can use the NOization procedure to reduce the depth of the variational circuit as the orbitals get rotated.

The specifics of our implementation are the following:

(1) We pick, as a set of generators $P$ to construct our gates $e^{i\theta P}$ from, the particle-number-preserving two-qubit operators $X_i X_j + Y_i Y_j$, the non particle-number-preserving two-qubit operators $X_i Y_j$, the density-density-like operators $Z_i Z_j$, and the single-qubit generators $X_i, Y_i$ and $Z_i$. All qubit pairs $(i,j)$ are allowed. 
    
(2) We get a reference state $|\psi_{\mathrm{ref}} \rangle $ by running VQE on a product circuit $\otimes RY(\theta_i, q_i)$ (rotation of qubit $q_i$ with an angle $\theta_i$ around the $y$ axis), as in \cite{Gyawali2022}. We do this rather than starting from some product state $\otimes X$ because the latter has vanishing gradients for the operators of the pool considered here.
    
(3) We run qubit-ADAPT VQE with our previously-defined pool, with a budget of 10 circuit growth steps. At step $n$, the operator $e^{i\theta_n P_n}$ yielding the highest-magnitude gradient at $\theta_n = 0$ over the variational state 
\begin{equation}
    |\psi(\bm{\theta_{n-1}^*}, 0)\rangle = \prod_{m=1}^{n-1}P_m(\theta_m^*)|\psi_{\mathrm{ref}} \rangle
\end{equation}   
is added to the circuit. The magnitude around $\theta_k=0$
of the gradient associated with generator $P_k$ can be shown \cite{grimsley_adaptive_2019} to be expressible as
\begin{equation}
    G_k = \left|\langle \psi(\bm{\theta_{n-1}^*}, 0)| [H, P_k] | \psi(\bm{\theta_{n-1}^*}, 0)\rangle \right|.
\end{equation}
All of the parameters of the circuit are then reoptimized within VQE with the COBYLA optimizer, except for the parameters defining the reference state.
    
(4) We perform orbital rotation with NOization on the last converged variational state.
    
(5) We repeat steps 2 to 4 several times, either until a budget of orbital rotations has been exhausted or until no further improvement is brought about by the orbital updates. Here we repeated the adaptive procedure five times. 

In both schemes, illustrated on Fig.~\ref{fig:method}, the orbital rotation at step $k$ is determined by the 1-RDM $D^{(k)}$ of the current approximate ground state $\ket{\psi^{(k)}(\vec{\theta^*})}$: it is the basis change which diagonalizes this matrix. The VQE runs consist in 'circuit learning' in that they either simply optimize the variational parameters of a fixed ansatz, or determine both the ansatz circuit and its optimal parameters. The orbital rotation steps can be understood as 'representation learning' steps. 

We note that Ref.~\cite{Fitzpatrick2022} uses a similar combination of adaptive VQE with orbital optimization, where the latter is however guided by the two-particle reduced density matrix (2-RDM).

In the next section, we present results with and without noise for the Hubbard model with 2 and 4 sites, with onsite interactions $U=0$ and $U=1$, namely in an uncorrelated regime ($U=0$), where the ground state is a Slater determinant, but also in a correlated regime ($U=1$).
We first document the expressiveness enhancement allowed by the NOization procedure when using a fixed circuit structure.

\begin{figure}
\begin{centering}
\includegraphics[width=0.85\columnwidth]{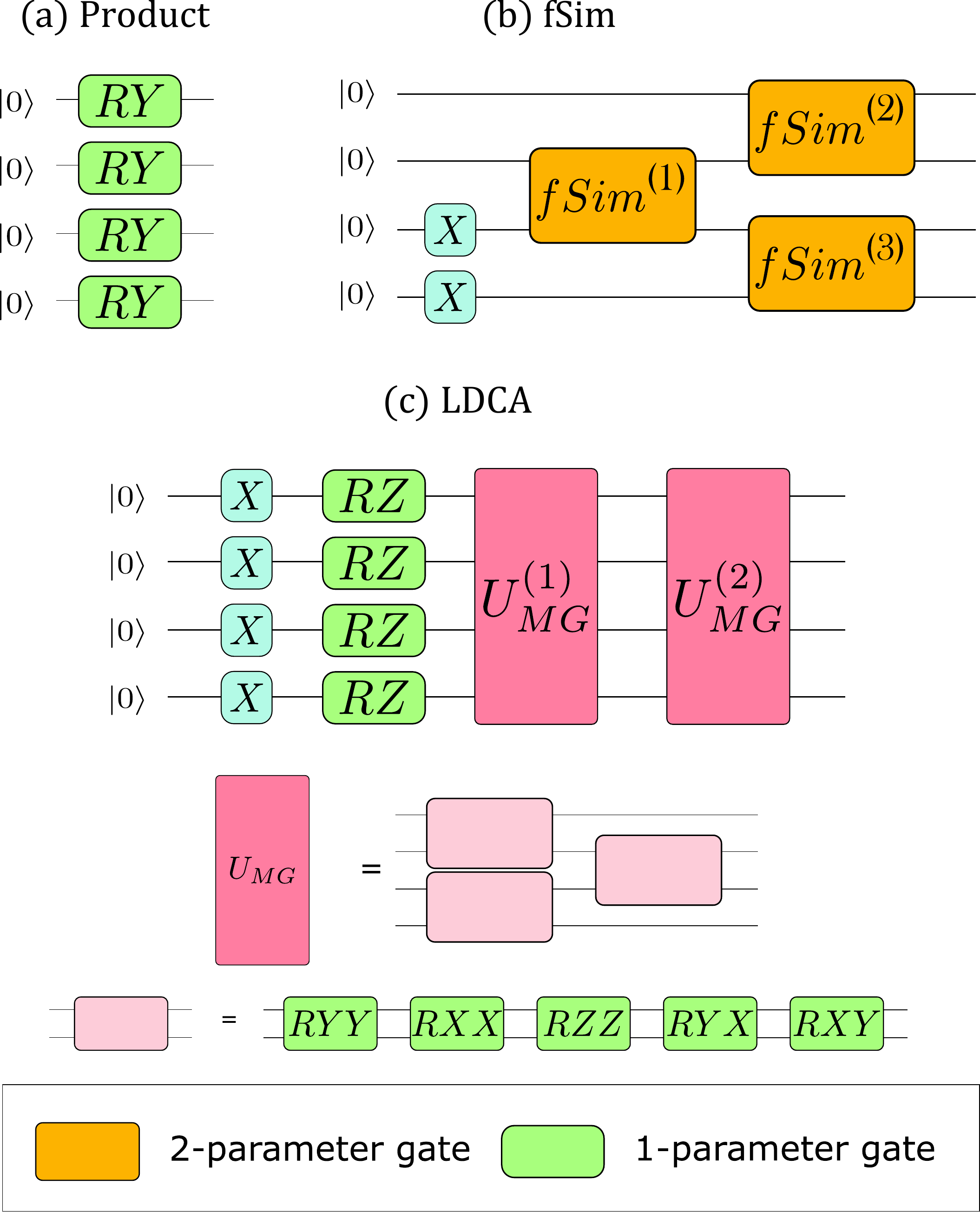}
\par\end{centering}
\caption{\emph{Ansatz circuits}, illustrated in the 4-qubit case (Hubbard dimer, $N=2$). \label{fig:circuits}}
\end{figure}

\section{Results}

\begin{figure*}
\centering
       \includegraphics[width=\textwidth]{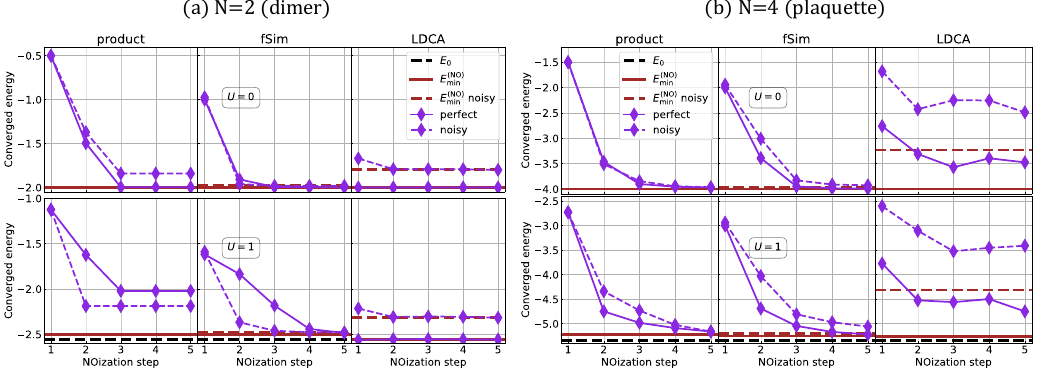}
    \caption{Converged VQE energies for the half-filled Hubbard model at each step of the natural-orbitalizing scheme for the three ansatz circuits of Fig.~\ref{fig:circuits}. $N=2$ (left panel) and $N=4$ (right panel) sites, with interaction $U=0$ (top row) and $U=1$ (bottom row). Noiseless and noisy (depolarizing noise) simulations.
    \label{fig:EE_final}}
\end{figure*}

In this section, we apply the NOization method, in its fixed-ansatz and adaptive versions, to the Hubbard model, defined by the Hamiltonian
\begin{equation}
    H = -t \sum \limits_{\langle i, j \rangle, \sigma}  (c^{\dagger}_{i\sigma}c_{j\sigma} + \mathrm{h.c.}) + U\sum \limits_i n_{i\uparrow} n_{i\downarrow} - \mu\sum \limits_i (n_{i\uparrow} + n_{i\downarrow}) 
    \label{eq:Hubbard}
\end{equation}
where $t$ and $U$ denote respectively the nearest-neighbor ($\langle i,j \rangle$) tunneling amplitude and on-site interaction of fermions of spin $\sigma=\uparrow,\downarrow$ on a lattice with sites $i=1\dots N$. We set the chemical potential $\mu=U/2$ to enforce half-filling of the ground state and fix the reference energy scale through $t=1$. Values of $U$ must thus be understood as values of the ratio $U/t$.
We will investigate the non-interacting regime $U=0$ and a correlated regime, $U=1$. The number of sites considered will be $N=2$ (Hubbard dimer) as well as $N=4$ in a square lattice geometry (Hubbard plaquette).

To transform $H$ into a sum of Pauli terms, we use the Jordan-Wigner transformation \cite{JordanWigner1928}:
\begin{subequations}
\begin{align}
    \label{eq:JW}
    & c^{\dagger}_p \rightarrow \hat{Z}_1 \otimes \hat{Z}_2 \otimes ... \otimes \hat{Z}_{p-1} \otimes \frac{1}{2} (\hat{X}_p - i \hat{Y}_p), \\
    & c_p \rightarrow \hat{Z}_1 \otimes \hat{Z}_2 \otimes ... \otimes \hat{Z}_{p-1} \otimes \frac{1}{2} (\hat{X}_p + i \hat{Y}_p),
\end{align}
\end{subequations}
which maps the $M=2N$ spin-orbitals $p = (i, \sigma)$ onto qubits.

Whenever considering the effect of decoherence, we add a depolarizing noise channel after each one-qubit gate:
\begin{equation}
    \mathcal{E}^{(1)}_{p_1}(\rho) = (1-p_1)\rho + \frac{p_1}{3}\left( X \rho X + Y \rho Y + Z \rho Z \right).
\end{equation}
After each two-qubit gate, likewise, we add a two-qubit channel $\mathcal{E}^{(2)}_{p_2}(\rho) = (\mathcal{E}_{p_2}^{(1)} \otimes \mathcal{E}_{p_2}^{(1)})(\rho)$.
We adjust the depolarizing probabilities $p_1$ and $p_2$ of the single and two-qubit depolarizing channels to match the error rates measured in randomized benchmarking experiments on current NISQ processors, using the relation \cite{magesan_characterizing_2012}
$p = 1-\left(1+\frac{1}{d} \right)\epsilon_\mathrm{RB}$,
with $d=2^n$ the dimension of the subspace that is acted on by the channel. For one and two-qubit depolarizing noise, this gives $p_1 = \frac{3}{2} \epsilon_\mathrm{RB}^{(1)}$ and $p_2 = 1-\sqrt{1-\frac{5}{4}\epsilon_\mathrm{RB}^{(2)}}$.

The specific RB values we choose are those referenced by Google for the Sycamore chip as of 2019 \cite{arute_quantum_2019}: $\epsilon_\mathrm{RB}^{(1)} = 0.16\%$ and $\epsilon_\mathrm{RB}^{(2)} = 0.6 \%$.

Unless stated otherwise, VQE runs are executed with a limit of 1000 COBYLA optimization steps. We consider several random initializations to avoid local minima (10 random initializations for subsection~\ref{subsec:expressiveness_enhancement}, 5 for subsection~\ref{subsec:tradeoff}).

\subsection{Expressiveness enhancement via natural orbitalization}\label{subsec:expressiveness_enhancement}

We investigate the NOization method for three fixed ansätze documented on Fig.~\ref{fig:circuits}:

(i) ``product'' consists in a product of $RY$ parametrized rotations. It produces only product (factorized) states, and is adapted only to non-interacting Hamiltonians expressed in their diagonal basis;

(ii) ``fSim'' consists in $X$ gates on half of the qubits to fix the number of excitations, followed by a routine made of excitation-number preserving Google's fSim gates \cite{Foxen2020, arute_hartree-fock_2020} 
(with one layer here), see also \cite{cade_strategies_2019};

(iii) ``LDCA'' is the low-depth circuit ansatz \cite{Dallaire-Demers2018} (with one cycle here). Despite its name, it contains a relatively large number of gates, which gives it a high expressiveness, but makes it very sensitive to decoherence.

\subsubsection{Results for 2 sites}

Figure~\ref{fig:EE_final}(a) illustrates the behavior of the NOization procedure for the three above ansätze, different correlation regimes and noise levels.

For $U=0$ and in the absence of noise, all circuits, including the ``product'' ansatz, are able to recover the exact
energy after a few steps (in the NO basis, for $U=0$, the GS is a
product state).
Only the LDCA circuit is capable of reaching the target state directly in the original basis due to its very high expressiveness.
It thus does not benefit from the NOization procedure. When noise is included, the NOization procedure for the product circuit is slightly impacted.
Looking at the results for the fSim circuit, which is also very shallow but exhibits noisy results similar to that of perfect results, it is probably an artefact of the optimization and should not be observed as runs are repeated.
For LDCA, we observe that the converged energy is greatly degraded compared with noise-free performances. This is attributable to a large gate count which increases its sensitivity to hardware noise. Nevertheless, NOization interestingly has the effect of slightly improving the LDCA energy in the noisy setting.

\begin{figure}
    \centering
    \includegraphics[width=\columnwidth]{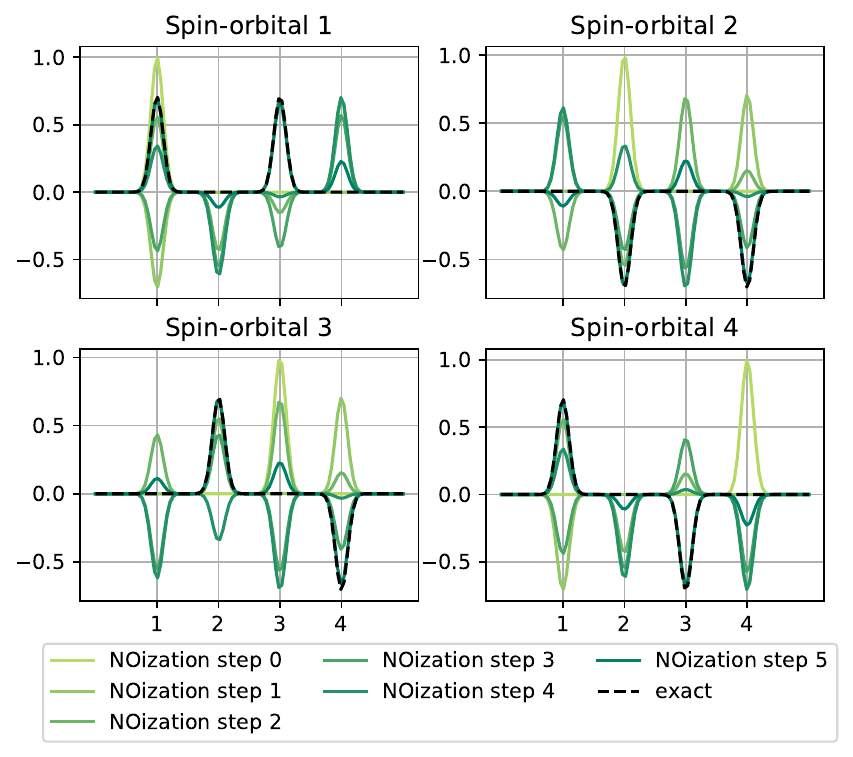}
    \caption{Evolution of the spin-orbitals along the NOization procedure for the Hubbard dimer ($N=2$), $U=1$, product ansatz circuit.}
    \label{fig:orbitals}
\end{figure}

For $U=1$, the ``product'' ansatz fails to reach the exact GS energy
because the true GS is entangled due to interactions. However, the energy gets lower as one rotates to a NO basis.
For illustration purposes we display on Fig.~\ref{fig:orbitals} the evolution of the spin-orbitals along the NOization procedure in this case. Originally, we start in the site-spin basis, which means we consider the single-fermion modes to be spanned by one Wannier orbital centered around each site per spin species. The natural-orbital basis associated with a two-fermion system is the so-called bonding-antibonding basis, which consists in  \textit{bonding} modes $\frac{1}{\sqrt{2}}(c^{\dagger}_{\sigma 1} + c^{\dagger}_{\sigma 2})$ and \textit{antibonding} modes $\frac{1}{\sqrt{2}}(c^{\dagger}_{\sigma 1} - c^{\dagger}_{\sigma 2})$. We observe that the procedure yields orbitals very close to this basis set, with a few spurious but small peaks.
This means that the energy bias is only attributable to the lack of expressiveness of the circuit even in NOs.
Note that this corresponds to the very particular case of a two-fermion system, where the set of ground state natural orbitals is independent of $U$.
Similarly the fSim circuit does not reach exactly the GS energy but gets very close as the orbitals get rotated.
LDCA retains its expressiveness and displays a behaviour very similar to the $U=0$ case.

The key point to be observed is that the NOization procedure is always found to yield improved performances: the converged VQE energy at NOization step $k+1$ is systematically observed to be lower than or equal to the energy at previous step $k$. The improvement goes as high as $75\%$ improvement for the product circuit at $U=0$. This substantiates the claim that the NOization procedure enables to break an expressiveness wall due to unadapted representation. 
It is remarkable that states which are greatly off-the-point such as the first converged VQE states for the product and fSim circuits provide orbital updates leading \textit{in fine} to the VQE energy they yield in the NO basis associated with the target ground state (materialized with brown lines on Fig.~\ref{fig:EE_final}(a)).

Raw VQE data is reported in Appendix \ref{sec:raw_vqe_traces}, in Fig.~\ref{fig:EE_2sites_perfect} and Fig.~\ref{fig:EE_2sites_noisy}.

\subsubsection{Results for 4 sites}

We now turn to the Hubbard plaquette ($N=4$) in order to make sure the good performances of the NOization procedure are not attributable to the special character of the two-fermion case. We now have four fermions and circuits with eight qubits. 

The results are summarized in Fig.~\ref{fig:EE_final} (b).
We can essentially make the same observations as for 2 sites. There is however a caveat: for the LDCA ansatz, convergence was not achieved after 1000 steps (see Fig.~\ref{fig:EE_4sites_perfect} and \ref{fig:EE_4sites_noisy}); hence, probably, the nonmonotonic
shape of LDCA curves in \ref{fig:EE_final} (b).

Raw VQE results without and with noise are given in Fig.~\ref{fig:EE_4sites_perfect} and \ref{fig:EE_4sites_noisy} of Appendix \ref{sec:raw_vqe_traces}.

\subsection{Trade-off between circuit sampling and gate noise}
\label{subsec:tradeoff}

\begin{figure*}
\centering
    \includegraphics[width=0.8\textwidth]{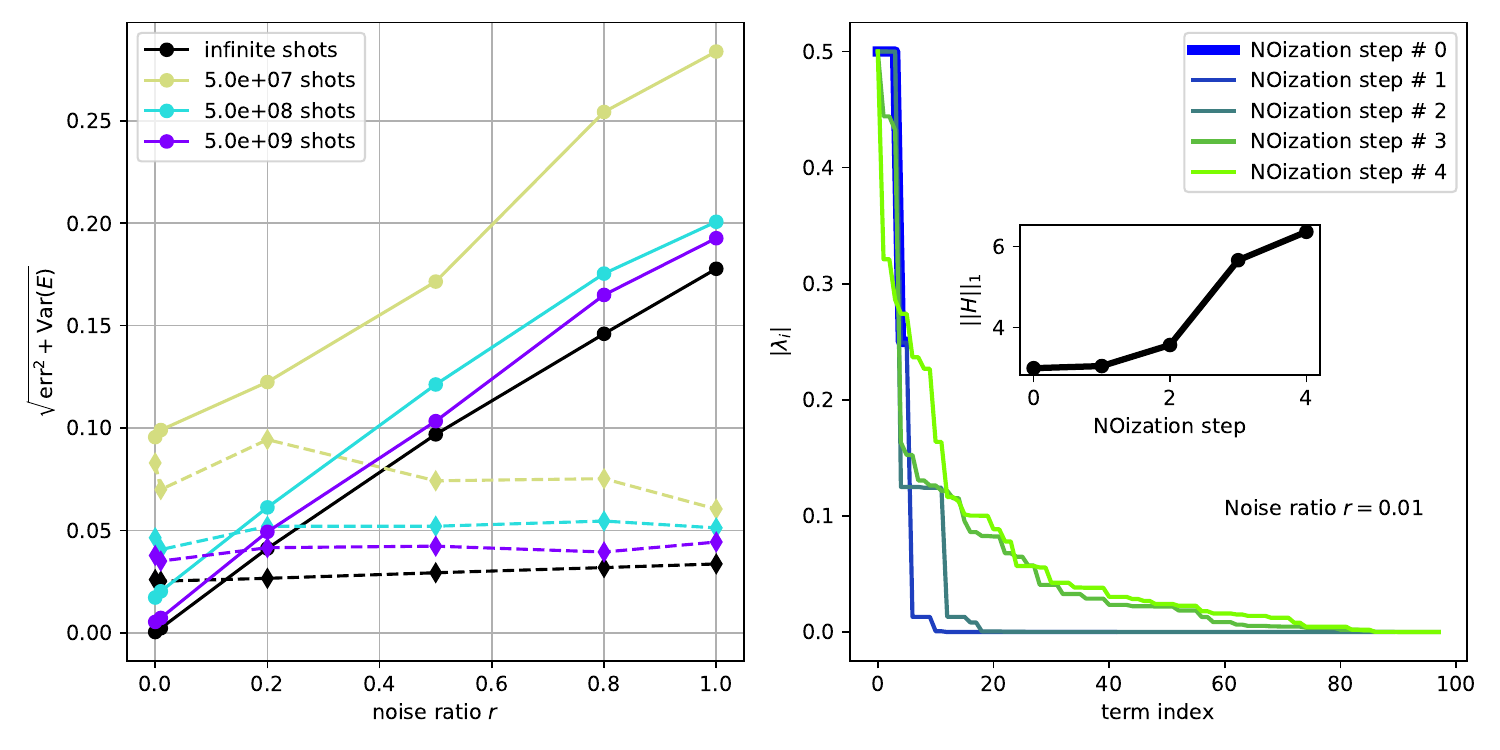}
\caption{\emph{Gate noise vs shot noise trade-off}. Fixed-ansatz natural-orbitalization strategy with the fSim circuit (dashed lines) compared to standard VQE with the LDCA ansatz (solid lines), $U=1$. \emph{Left}: total energy estimation error (bias and variance) as a function of the noise ratio $r$ for different total shot budgets $N_{\mathrm{shots}}$, averaged over 10 realizations. 
\emph{Right}: distribution of the Pauli coefficients $|\lambda_i|$ for the various NOization steps at $r=0.01$. Inset: one-norm as a function of the NOization step.
\label{fig:benchmark}}
\end{figure*}

In the previous sections, we focused on remedying the problem of quantum gates currently being considerably faulty and put forth an orbital-enhancing scheme based on natural orbitals.
As mentioned earlier, the price to pay for reduced requirements over circuit depth that come with this scheme is that the number of terms in the Hamiltonian may increase.
As a consequence, given a fixed number of shots for each energy evaluation, the statistical variance (or "shot noise") in the energy measurement may increase because each individual term may receive a smallest share of the shot budget.
Alternatively, if one allots a number of shots proportional to the relative strength of each term \cite{Rubin2018}, the statistical error is bounded by the one-norm of the Hamiltonian. Changing the orbital basis likely changes this one-norm, thereby possibly increasing shot noise.

In this section, we carry out a study on the specific case of the half-filled  Hubbard dimer at $U/t=1$ to investigate the trade-off between gate noise and shot noise.
We compare the effect of shot noise on standard VQE (with the LDCA ansatz) and on NOization strategy with the fSim ansatz (Fig.~\ref{fig:circuits}(b)).
Note that unlike the LDCA ansatz which does provide a means to reach the ground state energy $E_0=-2.56$ in the absence of noise, this ansatz was not found to reach energies lower than $E_{\mathrm{min}}^{\mathrm{(fSim)}} = -2.50$ even in noise-free simulation of VQE in natural orbitals. This means that below a certain level of noise, the LDCA ansatz will always provide better performances than the NOized fSim ansatz. 
We try here to determine what is this level of noise, and how it changes with the number of shots.

The depolarizing noise levels we consider are chosen relatively to current NISQ noise levels. As mentioned in the introduction to this section, we set the maximum noise level to correspond to that of Google's Sycamore chip \cite{arute_quantum_2019}.
We investigate fractions $r$ of this noise level in order to take into account conceivable future hardware improvements, namely we consider depolarizing QPUs with
\begin{equation}
    \epsilon_{\mathrm{RB}}^{(1)}(r) = r \times \epsilon_{\mathrm{RB}}^{(1)}, \;\;\;
    \epsilon_{\mathrm{RB}}^{(2)}(r) = r \times \epsilon_{\mathrm{RB}}^{(2)}.
\end{equation}

To observe the effect of shot noise, we consider different budgets for the total number of shots $n^\mathrm{tot}_{\mathrm{shots}}$ allocated to the task of estimating the ground state energy. At the most atomistic level this total budget is turned into a budget $n_{\mathrm{shots}}$ to evaluate the expectation value over the current variational circuit.
These $n_{\mathrm{shots}}$ shots are allocated proportionally to the absolute value of the term's coefficient, $n_{\mathrm{shots}}^{(i)} \propto |\lambda_i|$, a strategy that yields an upper bound $\left\Vert \tilde{H}(\bm{\theta'})\right\Vert_1 / \sqrt{n_{\mathrm{shots}}}$ \cite{wecker_towards_2015,rubin_application_2018} to the standard error on the mean.
For a single VQE procedure, we allow for up to $n_{\mathrm{iter}}=1000$ optimization steps (here, with the COBYLA minimizer).
When NOizing, the VQE procedure is repeated a number $K$ of times.
Also, to account for sensitivity of the VQE result to the initial parameters we repeat each VQE step a number $n_{\mathrm{repeats}}=5$ of times with random initialization, and keep the lowest-energy result.
All in all, each energy measurement of in the NOizing scheme is given a budget
\begin{equation}
    n_{\mathrm{shots}}^{\mathrm{(NOizing)}} = \frac{n^\mathrm{tot}_{\mathrm{shots}}}{n_{\mathrm{iter}} \times n_{\mathrm{repeats}} \times K}
\end{equation}
whereas in the direct LDCA scheme we allocate a number
\begin{equation}
    n_{\mathrm{shots}}^{\mathrm{(direct)}} = \frac{n^\mathrm{tot}_{\mathrm{shots}}}{ n_{\mathrm{iter}} \times n_{\mathrm{repeats}}}.
\end{equation}

Note that in the NOization procedure, compared to the straightforward one, we have additional 1-RDM measurements. We neglect their shot budget consumption (here, it represents $16 \times n_{\mathrm{shots}}^{\mathrm{(NOizing)}}$ additional shots per NOization step whereas a single VQE run consumes $n_{\mathrm{iter}}\times n_{\mathrm{shots}}^{\mathrm{(NOizing)}}$ with $n_{\mathrm{iter}} = 1000$).

The final energy error stems from a bias and a variance term
 \begin{align}
     \mathrm{err} &= \left\langle \left|\frac{E- E_0}{E_0} \right| \right\rangle,\\
     \mathrm{Var}(E) &= \left\langle \left(\frac{E - \langle E \rangle}{E_0} \right)^2 \right\rangle,
     \label{eq:err}
 \end{align}
with $E_0$ the exact ground state energy. The average here corresponds to repeating the whole VQE+NOization procedure 10 times. We will consider the mixed figure of merit $\sqrt{\mathrm{err}^2 + \mathrm{Var}(E)}$.
 
The results are summarized on Fig.~\ref{fig:benchmark}.
On the left panel, first of all, we observe that both the standard VQE (LDCA) and the NOization (fSim) approaches are  impacted by shot noise in a similar fashion.
As for the effect of depolarizing noise, we observe a roughly linear increase in $\sqrt{\mathrm{err}^2 + \mathrm{Var}(E)}$ with $r$ for the standard VQE (LDCA) approach, as opposed to virtually no impact on the NOization (fSim) method. 

To explain why shot noise does not impact NOization substantially more than it does VQE, we study the distribution of the Pauli weights for different NO steps in the right panel of Fig.~\ref{fig:benchmark}. 
In the original (site-spin) basis, the number of terms is very small (6 terms) with substantial weight for all terms.
As NOization proceeds, the number of terms increases (reaching a hundred terms). Yet, the distribution of the coefficients remains concentrated around a few sizable contributions, with a lot of small coefficients.
Because we allocate the number of shots proportionally to the weight, rotating to the NO basis thus does not substantially change the shot noise.
This can be checked more quantitatively by looking at the evolution of the one-norm (see inset), which only doubles when going from the original basis to the NO basis.

Another striking observation is that for the smallest shot count ($5\cdot 10^7$ shots, corresponding to $10,000$ shots per energy evaluation), NOization seems to outperform standard VQE for all decoherence levels.

\subsection{Circuit shortening with the natural orbitalizating adaptive VQE (NOA-VQE) approach}

\begin{figure*}
    \centering
    \includegraphics[width=0.8\textwidth]{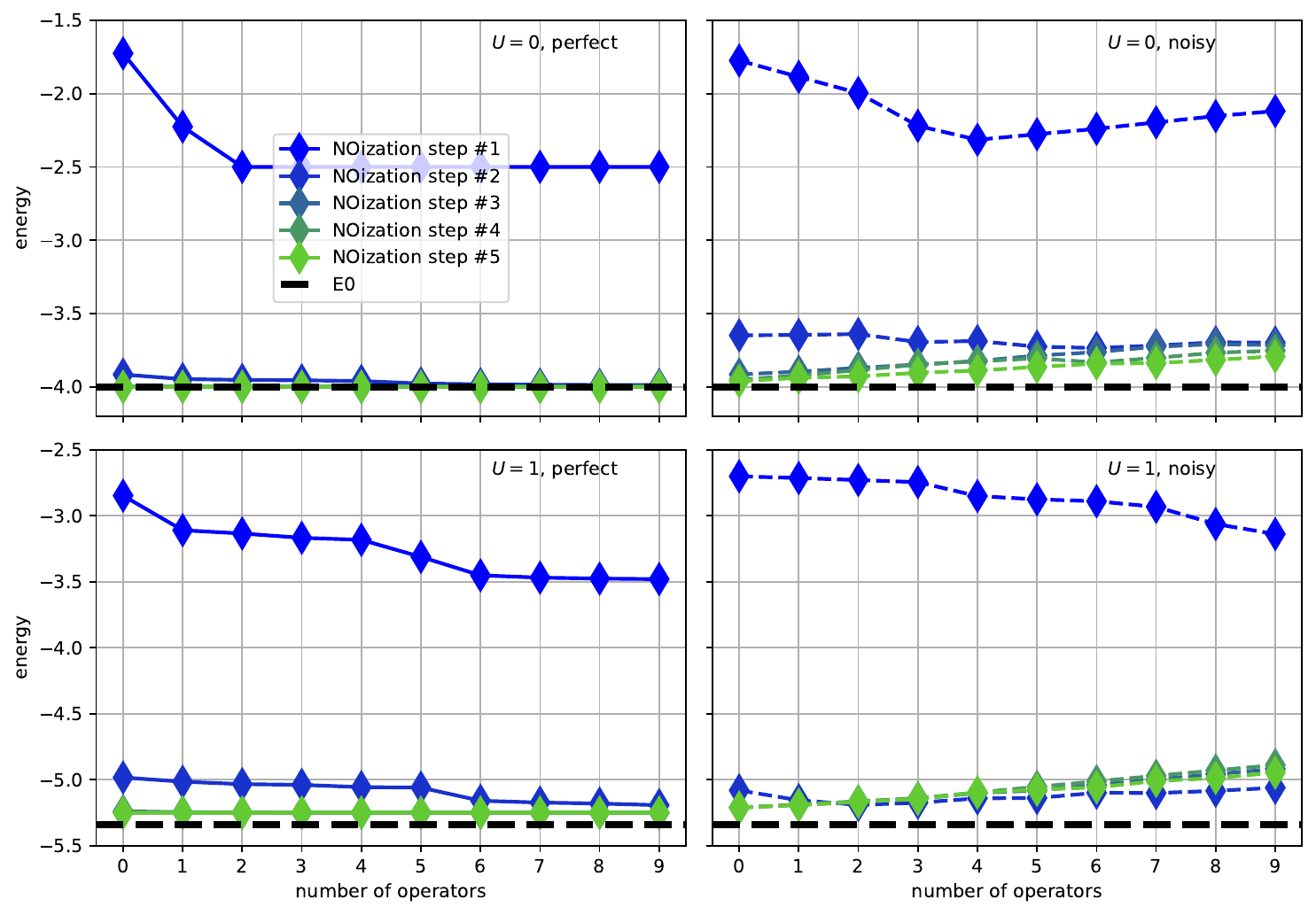}
    \caption{Adaptive natural orbitalization for the Hubbard plaquette ($N=4$). \emph{Upper panels:} $U=0$. \emph{Lower panels:} $U=1$. \emph{Left:} noiseless simulations. \emph{Right}: noisy simulations (depolarizing noise).}
    \label{fig:NOA_results}
\end{figure*} 

We now consider the adaptive version of the algorithm.
Results are presented for the more challenging four-site Hubbard model.
We still consider the $U=0$ and $U=1$ cases, with and without depolarizing noise.
Unlike the previous two sections, we do not decide on a circuit but rather construct it on the fly by selecting operators from a pool of our choice based on a gradient criterion, as detailed in Section \ref{sec:NOA-VQE}.
The pool corresponds to a QUBIT-ADAPT implementation \cite{tang_qubit-adapt-vqe_2021} with operators selected for their relevance regarding the physics of correlated fermionic states.
We consider a small budget of 10 operator additions so as to focus the study on the effect of orbital updates.

Results are presented on Fig.~\ref{fig:NOA_results}.
We observe a systematic improvement in terms of energy from one run to another, both in terms of the energy of the initial, reference state (as wall of $RY$ gates) as well as in terms of the final energy after the final ten-operator circuit is optimized.
In all settings, we observe that the first convergence in terms of NO steps is very quick: at the second step, already, the method reaches energies very close to the exact ground state energy.
In the $U=0$, noiseless case, the convergence to the exact ground state energy is perfect.
Going to a correlated case ($U=1$), we get a converged energy that is close, although not exactly equal to the exact ground state energy. We ascribe this small deviation to the quite simple pool we picked, the lack of a smart initialization, and the small number of ADAPT-VQE steps we chose.

In the presence of depolarizing noise, results are similar to the noiseless case, namely an improvement with NOization, with the important difference that adding more operators in the adaptive ansatz is seen to degrade the converged energy in some cases. This is because longer circuits, although possibly more expressive, are more sensitive to noise.

\section{Conclusion}

In this work, we proposed a method to reduce the required gate count for the task of variational state preparation of fermionic ground states.

More precisely, we put forth an orbital update technique that allows to tailor the single-particle basis to the target state and thus incorporate part of the correlations of the target state at the level of the Hamiltonian.
This contrasts with the usual paradigm in which one relies on the variational circuit to generate these correlations, as well as with recent proposals which carry out full-fledged  orbital optimization. The method consists in interleaving VQE runs with rotation of the single-particle modes to the natural-orbital basis of the current optimized VQE state. 

It was tested both within perfect state simulation (assuming a perfect QPU) and in a noisy setting, with a depolarizing model representative of recent devices. The systems to which the method was applied are instances of the Hubbard model with two and four sites, without ($U/t=0$) and in the presence of ($U/t=1$) correlations. 

Within usual VQE---that is, with a fixed circuit structure---we evidence increased expressiveness in the sense that the optimized energy gets lowered as the orbitals get updated in both cases.
Within an adaptive scheme, where the circuit is constructed on the fly, we observe that reduced depths are necessary to go below a given energy threshold as the orbitals get updated. The method is relevant in the presence of noise as it provides a way to resort to small variational circuits that are less sensitive to noise. This contrasts with deep circuits that perform well without noise such as the low-depth circuit ansatz or the unitary coupled cluster ansatz, as these are not compatible with current noise levels.  

One main drawback of the method is that orbital rotations do not preserve the number of terms of the Hamiltonian or its one-norm. As a consequence, the Hamiltonian observable to be measured onto circuits has more terms. This means that a careful analysis of resources, both in terms of qubit quality and in terms of shot budget, must be performed in order to assess the relevance of our method.
We show that at current and improved noise levels, for the ansätze we studied, NOization is still preferable over standard VQE.

Another drawback of our method concerns asymmetric models such as impurity models in which part of the qubits represent correlated impurities whereas the others represent uncorrelated 'bath' modes. In this context, if the different kinds of modes do mix---as is allowed in the absence of further restrictions to the method---then the correlation term does not concern the few impurity modes anymore, but is being spread onto all of the modes by orbital rotation. This prevents leveraging the different kind of physics different parts of the qubit register must display.
A way to avoid this would be to restrict the modes rotations to the bath qubits, as is done in classical settings, for instance in \cite{debertolis_few-body_2021}. 

Finally, one could think about other kinds of orbital updates. For instance, the basis that minimizes the seniority of the state (that is, the number of unpaired electrons) is associated with a lower informational content (the entropy of the probability distribution of the state across the different computational basis states) \cite{alcoba_configuration_2014}. From a physics point of view, the set of natural orbitals is only concerned with single-particle degrees of freedom, as it is defined from the 1-RDM. It would be interesting to identify a set of spin-orbitals which are tailored to a target state according to, also, 2-RDM information. 

\acknowledgments
We acknowledge useful discussions with Gaurav Gyawali.
This work is part of HQI initiative (www.hqi.fr) and is supported by France 2030 under the French National Research Agency award number “ANR-22-PNCQ-0002”.
The quantum circuit emulations were performed on the Eviden Qaptiva platform. 

\appendix

\setcounter{figure}{0}
\renewcommand\thefigure{A.\arabic{figure}} 

\section{Proof of the correlation-entropy minimization property of natural orbitals}
\label{app:proof_s_corr_min}
Let's introduce the so-called \textit{correlation entropy}, defined for any matrix $A$ with non-negative diagonal entries as

\begin{equation}
    S_{\mathrm{corr}}(A) \equiv - \sum_i A_{ii} \log{A_{ii}}
\end{equation}

From definition \eqref{eqn:def_1RDM}, such a quantity can be defined for $D$ since diagonal elements are occupation numbers associated to the single-particle orbitals defined by the set of creation operators $\{c^{\dagger}_i\}$. It provides a measure of the spread of the distribution onto the different orbitals of the $\mathrm{Tr}(D) \equiv n_{\mathrm{part}}$ particles $\ket{\psi}$ describes the state of. 

We aim at showing that the transformation to NO 
\begin{equation}
    D = UnU^{\dagger}
\end{equation}
yields a minimal correlation entropy 1-RDM (and thus, a maximally localized occupation numbers distribution in a sense that will be made clearer later):
\begin{equation}
    U = \argmin_{V \in U(M)}(S_{\mathrm{corr}}(V^{\dagger} D V))
\end{equation}
or, in other words, 
\begin{equation}
    S_{\mathrm{corr}}(D) \geq S_{\mathrm{corr}}(n)
\end{equation}

Since $n$ is diagonal we have that $S_{\mathrm{corr}}(n) = -\mathrm{Tr}(n\log n) \equiv S(n)$ where $S$ denotes the von-Neumann entropy. Since the von-Neumann entropy is invariant under unitary transformation, we also have $S(n) = S(D)$. Conversely, $S_{\mathrm{corr}}(D) = S(\mathrm{diag}(D))$ with $\mathrm{diag}(D)$ referring to the diagonal matrix obtained from $D$ by setting its off-diagonal elements to 0. $S(D)$ and $S(\mathrm{diag}(D))$ are related by the so-called \textit{relative entropy of coherence} \cite{baumgratz_quantifying_2014} $\mathcal{C}_{\mathrm{RE}}(D) \equiv S(\mathrm{diag}(D)) - S(D) = S_{\mathrm{corr}}(D) - S_{\mathrm{corr}}(n)$. From Ref. \cite{baumgratz_quantifying_2014} we can rewrite $\mathcal{C}_{\mathrm{RE}}(D^1)$ as the minimum of the relative entropy $S(D || \delta ) \equiv \mathrm{Tr}(D \log D) - \mathrm{Tr}(D \log \delta)$ with regards to incoherent (diagonal) matrices $\delta$. Since this quantity is known to be always positive (see eg \cite{wehrl_general_1978}), we have the result.

\section{Raw VQE traces}\label{sec:raw_vqe_traces}
\begin{figure*}
\centering
     \begin{subfigure}[t]{0.7\textwidth}
         \centering
         \includegraphics[width=\textwidth]{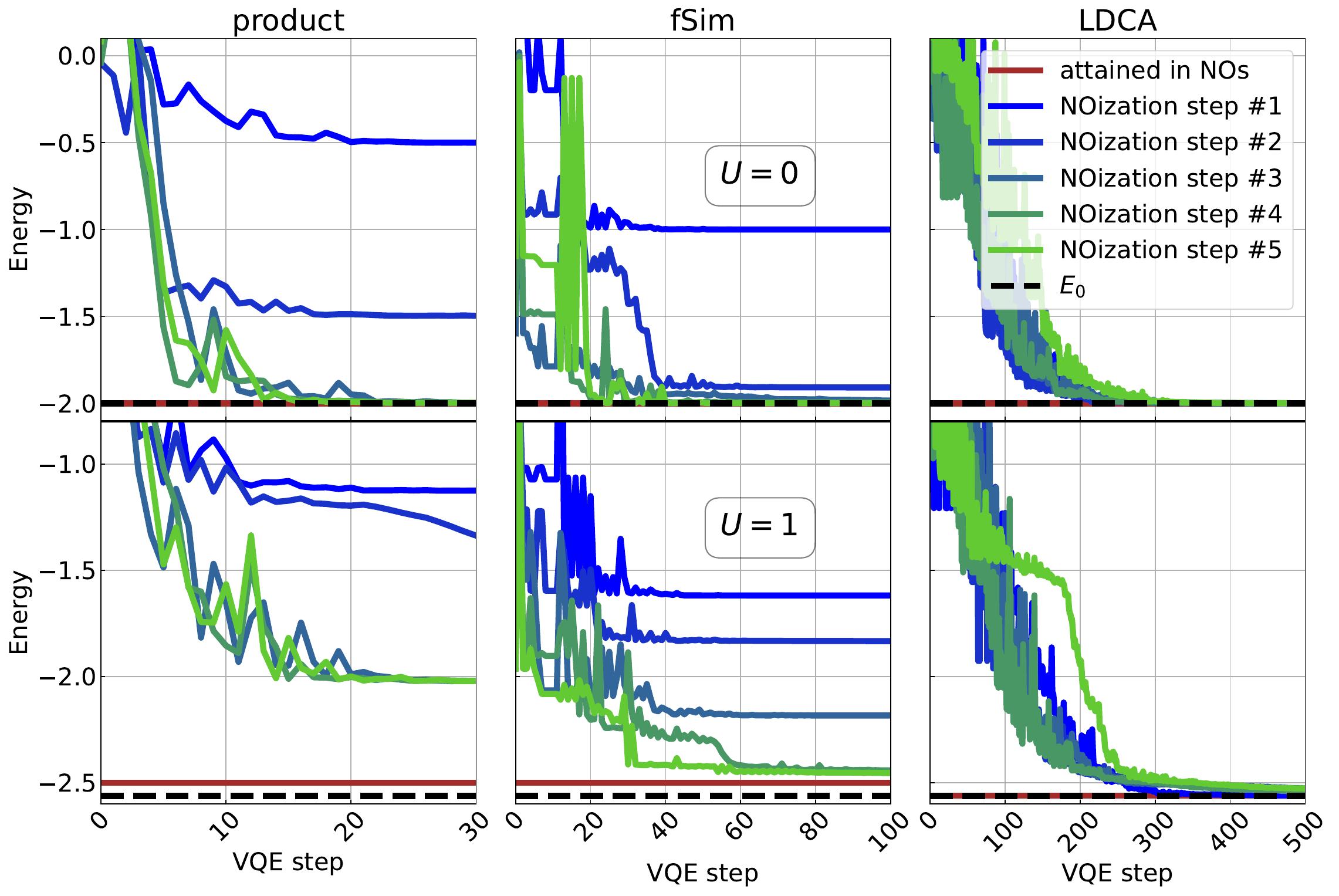}
         \caption{}
         \label{fig:EE_2sites_perfect}
    \end{subfigure}
     \\
     \begin{subfigure}[t]{0.7\textwidth}
         \centering
         \includegraphics[width=\textwidth]{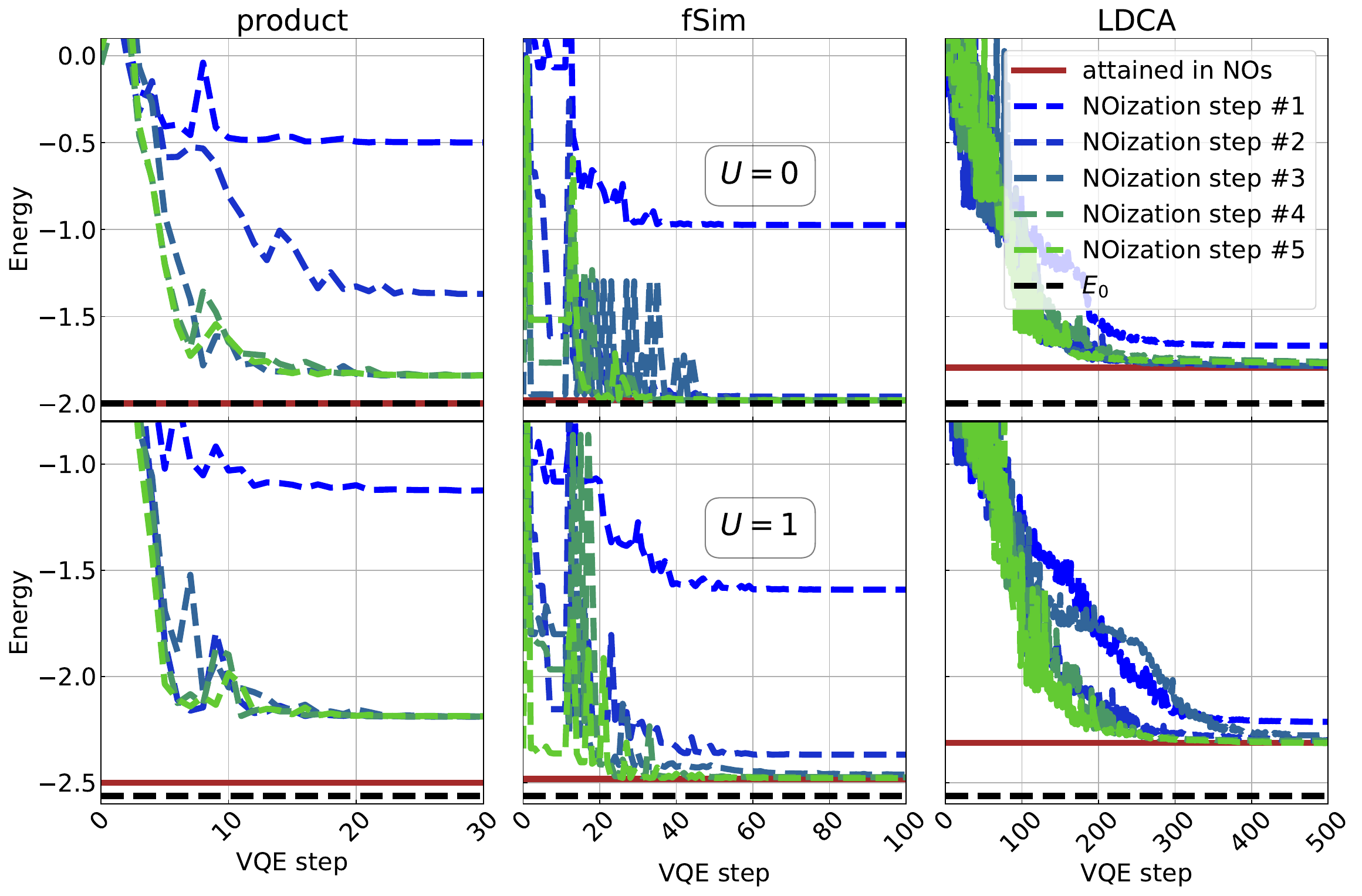}
        \caption{}
        \label{fig:EE_2sites_noisy}
    \end{subfigure}
         \caption{Raw VQE optimization traces at each step of the natural-orbitalizing scheme for the three ansatz circuits of Fig.~\ref{fig:circuits}. The method is applied here to the Hubbard model with $N=2$ sites, at half-filling and with Coulomb interaction $U=0$ and $U=1$. We consider both noise-free state preparation assuming a perfect QPU (upper panel \ref{fig:EE_2sites_perfect}) and noisy preparation with a depolarizing QPU model (lower panel \ref{fig:EE_2sites_noisy}) with noise rates representative of current devices.}
    \label{fig:EE_results_2sites}
\end{figure*}

\begin{figure*}
    \centering
    \begin{subfigure}[t]{0.7\textwidth}
         \centering
         \includegraphics[width=\textwidth]{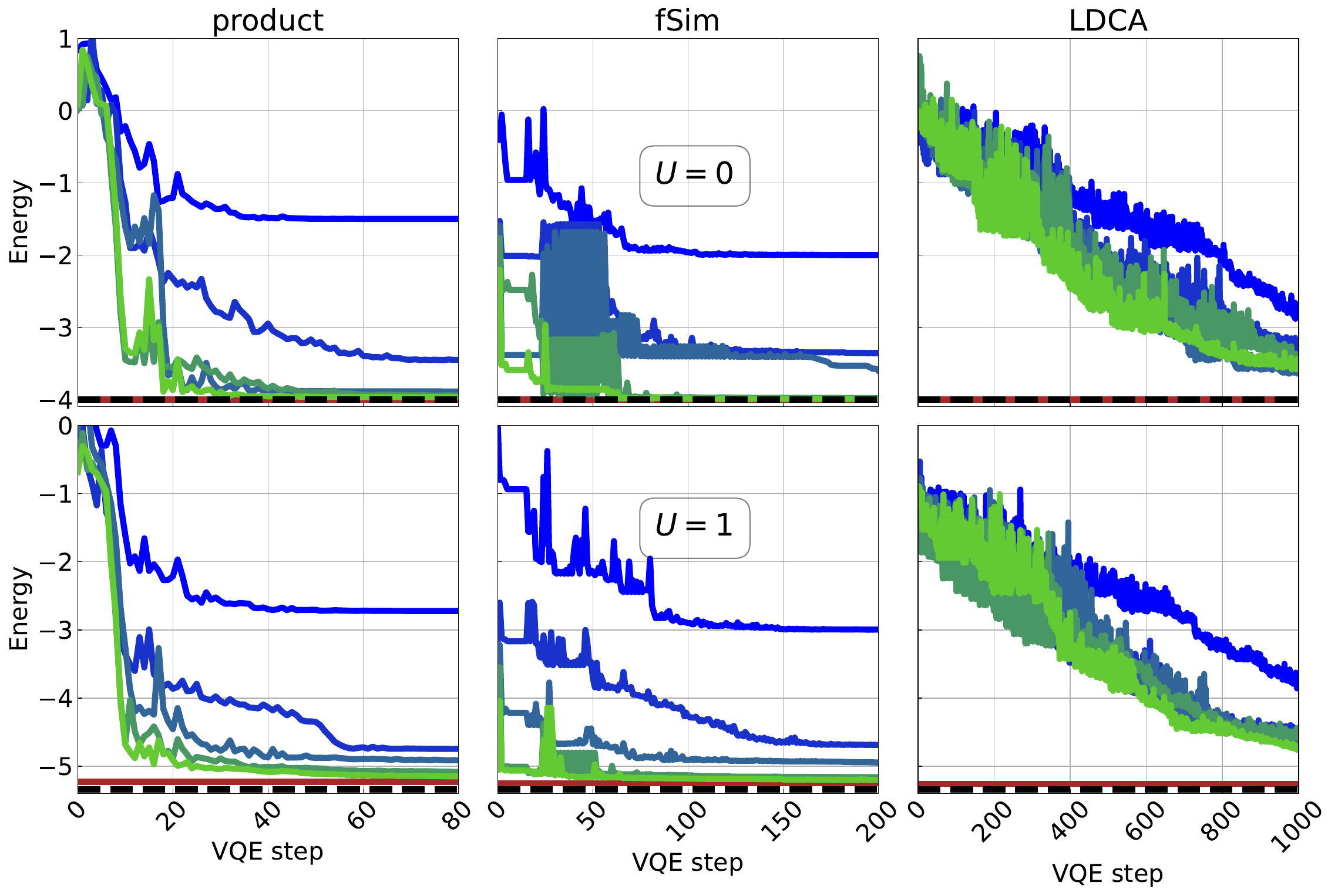}
          \caption{}
         \label{fig:EE_4sites_perfect}
    \end{subfigure}
    \\
    \begin{subfigure}[t]{0.7\textwidth}
         \centering
         \includegraphics[width=\textwidth]{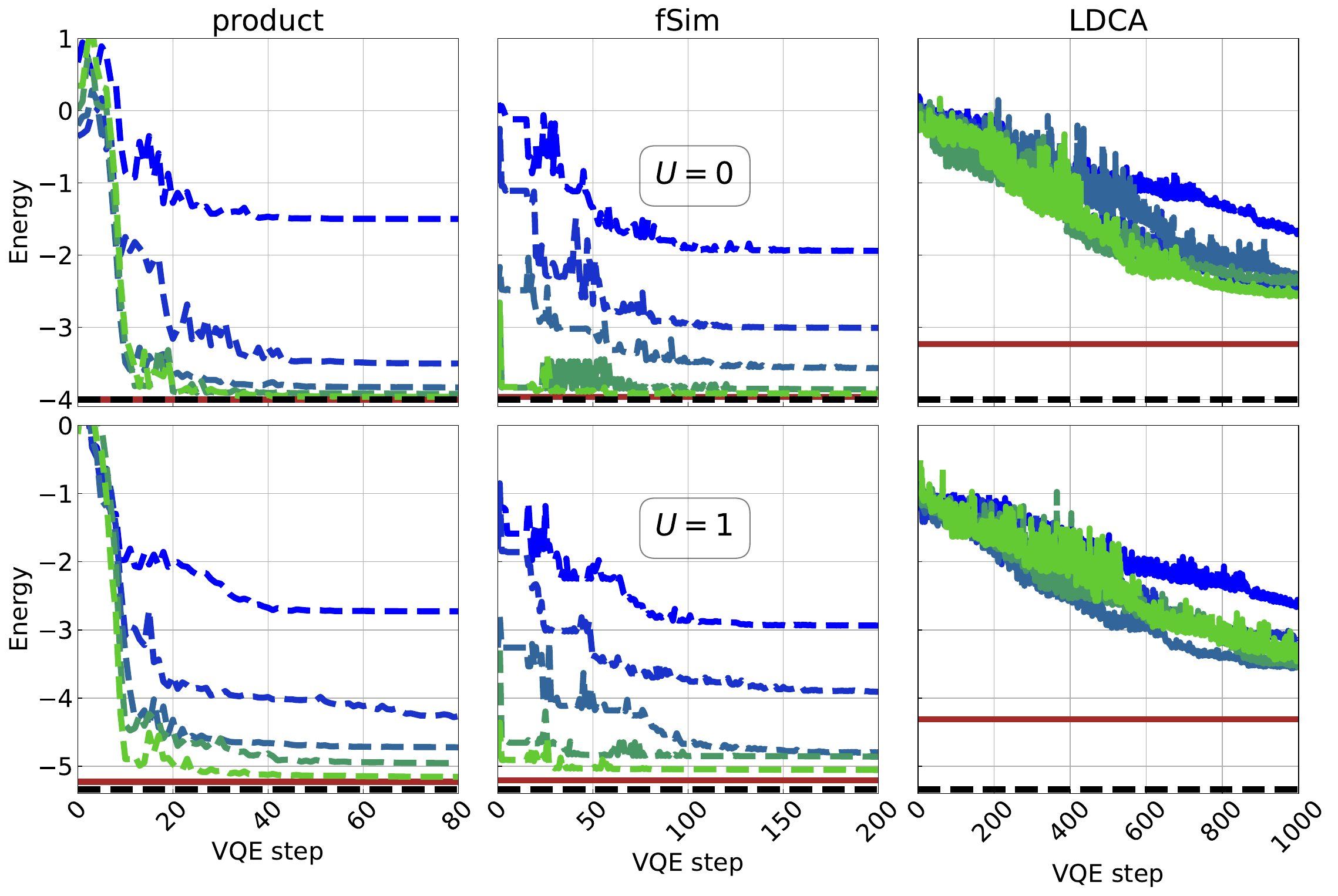}
         \caption{}
         \label{fig:EE_4sites_noisy}
    \end{subfigure}
    \caption{Same as Fig.~\ref{fig:EE_results_2sites}, with $N=4$ sites.}
    \label{fig:EE_results_4sites}
\end{figure*}

We report on Fig.~\ref{fig:EE_results_2sites} and Fig.~\ref{fig:EE_results_4sites} raw data for VQE results corresponding the different VQE optimizations along the NOization procedure for the Hubbard model with respectively two and four sites, at $U=0$ and $U=1$, with and without depolarizing noise.

\bibliographystyle{apsrev4-2}


%

\end{document}